\documentclass[fleqn,usenatbib]{mnras}

\usepackage{newtxtext,newtxmath}

\usepackage[T1]{fontenc}

\DeclareRobustCommand{\VAN}[3]{#2}
\let\VANthebibliography\thebibliography
\def\thebibliography{\DeclareRobustCommand{\VAN}[3]{##3}\VANthebibliography}

\usepackage{graphicx}	
\usepackage{amsmath}	
\usepackage{array}
\usepackage{multirow}
\usepackage[dvipsnames]{xcolor}

\title[GR mass-to-distance ratio of megamaser black holes]{A general relativistic 
mass-to-distance ratio for a set of megamaser AGN black holes}

\author[D. Villaraos et al.]{
D. Villaraos,$^{1}$\thanks{E-mail: deborahv@ifuap.buap.mx}
A. Herrera-Aguilar$^{1}$,
U. Nucamendi$^{2}$,
G. Gonz\'alez-Ju\'arez$^{3}$
and R. Lizardo-Castro$^{1}$
\\
$^{1}$Instituto de F\'\i{}sica, Benem\'{e}rita Universidad Aut\'{o}noma de Puebla,
Edificio IF-1, Ciudad Universitaria, CP 72570, Puebla, Puebla, M\'exico.\\
$^{2}$Instituto de F\'\i{}sica y Matem\'aticas, Universidad Michoacana de San Nicol\'{a}s de Hidalgo,
Edificio C–3, Ciudad Universitaria, CP 58040, Morelia, Michoac\'an, \\ M\'exico.\\
$^{3}$Facultad de Ciencias F\'\i{}sico Matem\'aticas, Benem\'{e}rita Universidad Aut\'{o}noma de Puebla,
Ciudad Universitaria, CP 72570, Puebla, Puebla, M\'exico.\\
}

\date{Accepted 2022 October 13. Received 2022 October 10; in original form 2022 July 12}

\pubyear{2022}

\begin{document}
\label{firstpage}
\pagerange{\pageref{firstpage}--\pageref{lastpage}}
\maketitle

\begin{abstract}
In this work we perform a Bayesian statistical fit to estimate the mass-to-distance ratio and the recessional redshift of 10 different black holes hosted at the centre of active galactic nuclei, namely the galaxies NGC 5765b, NGC 6323, UGC 3789, CGCG 074-064, ESO 558-G009, NGC 2960, NGC 6264, NGC 4388, J0437+2456 and NGC 2273. Our general relativistic method makes use of the positions in the sky and frequency shift observations of water megamasers circularly orbiting the central black hole on their accretion disks. This approach also allows us to quantify the gravitational redshift which is not considered in a Newtonian analysis. The gravitational redshift of the megamasers closest to the black hole is found to be within the range 1-6 km/s. The order of the fitted black hole masses corresponds to supermassive black holes and lies on the range $10^6 - 10^7 M_{\sun}$. 

\end{abstract}




\section{Introduction}

In the last decade there has been considerable observational evidence of the existence of black holes (BHs). From the detection of gravitational waves caused by the collision of two BHs carried out in 2015 by the LIGO and Virgo collaborations \citep{Ligo2016,LIGO2019} to the newly obtained images of BH shadows in M87 and our galaxy by the \citet{EHTI,EHTIV,EHTVI,EHTSGRI,EHTSGRII}. The existence of the latter BH (SgrA$^*$) was proposed after the determination of the mass distribution in the central parsec of the Milky Way and the observation of stars orbiting around a gravitational centre \citep{Genzel87,Genzel96,Ghez98}. Moreover, by analysing the kinematics of these stars it has been possible to determine some of the BH properties \citep{Ghez05,Ghez08,Gillessen09,Genzel10}.

H$_2$O megamasers consist of clouds of water vapour emitting at 22GHz by stimulated emission and are extremely luminous $L > 100L_{\sun}$; according to the model proposed by \citet{Claussen86}, megamasers arise in high-density circumnuclear disks with an active galactic nucleus (AGN) as the energy source for stimulated emission. In \citealt{Miyoshi95} and in \citealt{Herrnstein99} the authors use a Keplerian rotation curve to fit the maser trajectories in the BH accretion disk of NGC 4258. This megamaser system has been thoroughly studied, and it represents the archetypal model for megamaser configurations, showing a thin annular disk viewed edge-on. In particular, three groups of maser features are observed in the sky: the masers on the extremes of the disk (where they show a redshift and a blueshift given the rotation of the disk) and the central maser features \citep{KYLo05}. Henceforth, the former two maser groups will be called highly frequency-shifted masers and the masers located in the line of sight (LOS; the line that connects the observer with the BH) will be referred to as systemic masers.

Very Long Baseline Interferometry (VLBI) has been widely used to observe megamaser systems since it provides the submilliarcsecond resolution needed to map the sub-parsec maser disks \citep{Herrnstein99}. The Megamaser Cosmology Project\footnote{The Megamaser Cosmology Project is a National Radio Astronomy Observatory (NRAO) key project that seeks to measure the Hubble constant using megamaser observations of galaxies located on the Hubble flow. More information in: \url{https://safe.nrao.edu/wiki/bin/view/Main/MegamaserCosmologyProject}} (MCP) has mapped over 20 megamaser galaxies (e.g., \citealt{MCPIII,MCPIX,MCPXII}) using VLBI techniques with the Very Large Baseline Array (VLBA)\footnote{ The VLBA is operated by Associated Universities, Inc., under a cooperative agreement with the National Science Foundation which is a facility of the NRAO.}, the Radio Telescope Effelsberg (ET), the Robert C. Byrd Green Bank Telescope (GBT) and the Karl G. Jansky Very Large Array (VLA). In this work we chose 10 galaxies observed by the MCP to test our general relativistic model.

In general, Keplerian models with relativistic corrections are used to analyse BH systems (e.g., \citealt{Herrnstein05,Humphreys13,MCPIV}) and render results in good agreement with the observations. However, being the stars or, in this case, the megamasers in the gravitational field of a BH, general relativistic effects affect these astronomical objects and can, in principle, be detected. In fact, the \citet{Grav18,Grav20} was able to measure the gravitational redshift as well as the Schwarzschild precession of the S2 star that orbits SgrA$^*$ by using a parametrized post-Newtonian approach. Motivated by the possible detection of general relativistic effects in these systems, \citet{Herrera15} construct a full general relativistic model for a test particle that emits photons moving on a rotating axially symmetric space-time and further find expressions that link the mass of the BH with observable quantities, i.e. the redshift of the emitted photons \citep{Banerjee22}. In \citet{Nucamendi21} the authors use the mentioned model considering the Schwarzschild metric to estimate the BH parameters and quantify the corresponding gravitational redshift in the megamaser galaxy NGC 4258, while  \citet{Artemisa22} fit the BH parameters of the gigamaser TXS 2226–184, being the first BH mass estimation using gigamaser observations. 

Following this general relativistic approach, in this work we fit the mass-to-distance ratio $M/D$ and the recession velocity of the BHs hosted at the center of the galaxies NGC 5765b, NGC 6323, UGC 3789, CGCGC 074-064, ESO 558-G009, NGC 2960, NGC 6264, NGC 4388, J0437+2456 and NGC 2273 using the observations of the megamaser positions in the sky as well as their frequency shifts. The paper is organised as follows: In Sec. \ref{sec:GR} we present the theoretical general relativistic model for a massive test particle that emits photons and is circularly orbiting a Schwarzschild BH. In Sec. \ref{sec:obs} we briefly review the observations of the megamaser systems as well as previous estimates performed in other works. In Sec. \ref{sec:bayes} we explain the Bayesian statistical model and present the priors and the posteriors obtained after the fit for each galaxy. Finally, in Sec. \ref{sec:conclusions} we discuss the obtained results.

\section{General Relativistic Model} 
\label{sec:GR}

We start by considering a Schwarzschild BH metric in natural units:

\begin{align}
    ds^2 &=\frac{dr^2}{f} + r^2(d\theta ^2 + \sin ^2 \theta d\varphi^2) -fdt^2, & f&=1-\frac{2m}{r} .
\end{align}

A massive test particle is considered to be moving around the BH in circular orbits on the equatorial plane ($\theta = \pi /2 $), with 4-velocity $U^\mu  = (U^t, U^r, U^\theta, U^\varphi)$ normalized to unity $U^\mu U_\mu = -1$. 
The non-zero components of the 4-velocity for a particle moving under these conditions are

\begin{align}
    U^t(r, \pi/2)&= \sqrt{\frac{r}{r-3m}} ,  & U^\varphi (r, \pi/2)&= \pm \sqrt{\frac{m}{r-3m}}\frac{1}{r},
\end{align}
where the $\pm$ signs indicate the particle's direction of motion with respect to the observer. The test particle emits photons with 4-momentum $k^\mu = (k^t, k^r, k^\theta, k^\varphi)$ which move along null geodesics ($k^\mu k_\mu = 0$) towards the observer with frequency $\omega$. The Schwarzschild redshift $(_1)$ and blueshift $(_2)$ of the emitted photons are

\begin{align}
    1+ Z_{Schw 1,2} \equiv \frac{\omega_e}{\omega_d} = \frac{(U^t -b_{\mp}U^\varphi)|_e}{(U^t -b_{\mp}U^\varphi)|_d},
\end{align}
where the subscripts $(_e)$ and $(_d)$ indicate the measured frequency at the emission and detection points respectively.

The light bending parameter (or impact parameter) is, by definition, $b = L_\gamma / E_\gamma$, where $L_\gamma$ and $E_\gamma$ are the angular momentum and total energy of the emitted photons. Both quantities are conserved throughout the entire trajectory of the photons. The maximum value of the light bending parameter is reached at the extremes of the midline where $k^r =0$, and reads
\begin{align}
    b_\mp = \mp \sqrt{\frac{r^3}{r-2m}}.
\end{align}

We can consider the observer to be static $U^\mu|_d = (1,0,0,0)$ since it is located far away from the emitting source ($r_d \rightarrow \infty$). Then, the redshift expression is

\begin{align}
    1 +  Z_{Schw 1,2} = \sqrt{\frac{r_e}{r_e-3m}} \pm \sqrt{\frac{mr_e}{(r_e - 2m)(r_e - 3m)}}.
\end{align}

When $b = 0$, there is no radial component of the velocity and $k^{\varphi}=0$, therefore we obtain the redshift caused solely by the gravity of the BH, namely, the gravitational redshift

\begin{equation}
    1 + Z_{grav} = \sqrt{\frac{r_e}{r_e-3m}}.
    \label{eq:zgrav}
\end{equation}

Then, the kinematic redshift/blueshift at the extremes of the midline will be 

\begin{equation}
    Z_{kin\pm} =  Z_{Schw 1,2} -  Z_{grav} = \pm \sqrt{\frac{mr_e}{(r_e - 2m)(r_e - 3m)}},
\end{equation}
where the $_\pm$ subscripts correspond to the receding or approaching (redshift or blueshift) motion of the particle relative to the observer.

Moreover, we consider the cosmological redshift which is related with the location of the galaxy that hosts the BH into the Hubble flow i.e. the redshift due to the expansion of the Universe, $Z_{cosm}$.

Additionally, we compose this redshift with the frequency shift related with the peculiar motion of the galaxy, $Z_{pec}$, resulting in a total recessional redshift $Z_{rec}$ \citep{Davis14}
\begin{align}
    1+ Z_{rec} = (1+Z_{boost})(1+Z_{cosm}).
    \label{eq:zrec composition}
\end{align}
The peculiar redshift relates to a special relativistic boost
\begin{align}
    1+ Z_{boost} &= \frac{1+\beta\cos\alpha}{\sqrt{1-\beta^2}},
    \label{eq:zboost}
\end{align}
where $\beta \equiv v_{pec}/c$ and $v_{pec}=cZ_{pec}$ is the peculiar velocity of the galaxy. This peculiar redshift accounts for the velocity of the galaxy due to the gravitational interactions with its local group. The term $v_{pec}\cos\alpha$ accounts for the radial motion of the galaxy (we shall consider that $\alpha =0$ for simplicity).

On the other hand, the cosmological redshift depends on the model chosen for the expansion of the Universe. However, the metric we are using in order to compute the total redshift as a first approximation is static and provides no information about this expansion. The expression for the peculiar redshift does not depend on the metric itself, which leads to a degeneracy of the cosmological and the peculiar redshifts. Therefore in this work, for most of our galaxies, we will estimate the total recessional redshift in order to avoid this degeneracy.

Finally, the expression for the total redshift composing the Schwarzschild redshift and the recessional redshift reads as follows
\begin{align}
    1 + Z_{tot 1,2} = (1 + Z_{Schw 1,2})(1+ Z_{rec}).
    \label{eq:ztot}
\end{align}

\section{Observational data and previous estimates}
\label{sec:obs}

The observed astrophysical systems consist of thin disks of water masers geodesically orbiting around a BH in circular orbits viewed edge-on from Earth. In order to perform the statistical fit for each system, we use the observational data of the position and frequency shift of each maser feature obtained using VLBI techniques within the framework of the MCP.

For NGC 2273, NGC 2960, NGC 4388, NGC 6264 and NGC 6323 we use the data taken from 2005 to 2009 with the VLBA, GBT, and ET reported in \citet{MCPIII}; for UGC 3789 we use data obtained from 2006 to 2009 by the VLBA, GBT and ET in \citet{MCPIV}; for NGC 5765b we use data from 2012 to 2014 with the VLBA, GBT, ET and VLA in \citet{MCPVIII}; for ESO 558-G009 and J0437+2456 the data from 2010 to 2014 with VLBA, GBT, ET and VLA reported in \citet{MCPIX}; and for CGCG 074-064 data from 2015 to 2017 with VLBA, GBT and ET in \citet{MCPXI}.

For most of these systems, the position of the maser features was measured with respect to the brightest maser feature, which was not necessarily near the center of the disk. For this reason, we changed the origin to the geometric center formed by the LOS maser features and then rotated the disk by the position angle found by the MPC articles so that the maser disk lies horizontally on the equatorial plane.

Previous estimates performed by other works of the mass, distance, velocity and inclination angle of each megamaser system are listed in Table \ref{tab:prev}. Moreover, we calculate their mass-to-distance ratio to make a more direct comparison with our results, where we propagate the errors according to the following relation $\sigma _{M/D}^2 = (\sigma _{M}/D)^2 + (\sigma _{D} M/D^2)^2$, which are also shown in Table \ref{tab:prev}.

It is necessary to mention that in previous works related to the MCP, the authors use a Keplerian model with relativistic corrections to perform their estimates (see, for example, \citealt{MCPXI}). In such a model, they compose a Doppler redshift $z_D$  encoding the motion of the particle with a gravitational $z_g$ and a recessional $z_0$ ones for a total observational redshift $1 + z = (1+z_D)(1+z_g)(1+z_0)$. The gravitational redshift is given by $z_g = \left(1 - 2M/r_e\right)^{-1/2}$, where a massive static particle emitting photons is considered to be in the gravitational field of a Schwarzschild BH. This expression differs from Eq. (\ref{eq:zgrav}), since we consider a massive particle emitting photons and circularly orbiting the same spacetime.

\begin{table*}
\begin{center}
\caption{Statistical fits reported in previous works. Column 1: Name of the source. Column 2: Source position (J2000). All the source positions are taken from the corresponding references. Column 3: Bayesian fit for the BH mass. Column 4: Fitted distance to the source. Column 5: Mass to distance ratio calculation. We do not propagate the distance error on $M/D$ for the galaxies that do not have a reported error on $D$. Column 6: LOS velocity of the sources with respect to the Local Standard of Rest (LRS) with the optical definition of the redshift. The recession velocities are estimated by a Bayesian fit. Column 7: Bayesian fit for the inclination angle $\theta_0$ of the maser disk measured from the polar axis. For NGC 4388 the authors do not provide an inclination angle because of the lack of systemic maser features. Column 8: References for the reported data and statistical fits. All the aforementioned articles were published within the framework of the MCP where the authors use a Keplerian model supplemented by relativistic corrections. The superindex $^a$ indicates that errors were not reported.}
\begin{tabular}{ c c c c c c c c c }
Source & Position & Mass & Distance &  $M/D$ & Recessional & Inclination angle & Reference \\
& R.A. ($^h:^m :^s$) & $M$ & $D$ & & velocity & &\\
& Decl. ($\degr: ^{'} : ^{''}$) & ($ 10^7 M_{\odot}$) & (Mpc)  & ($ 10^5 M_{\odot}$)/Mpc & (km/s) & (°)& 
\\ \hline \hline 
\multirow{2}{*}{NGC 5765b} & 14:50:51.51950 & \multirow{2}{*}{4.55 $\pm$ 0.31} & \multirow{2}{*}{126.3 $\pm$ 11.6} & \multirow{2}{*}{3.602 $\pm$ 0.411} & \multirow{2}{*}{8334.60 $\pm$ 1.13} & \multirow{2}{*}{94.5 $\pm$ 0.25} & \multirow{2}{*}{\cite{MCPVIII}} \\
& 05:06:52.2502 & & & & & &\\
\hline
\multirow{2}{*}{NGC 6323} & 17:13:18.03991 & \multirow{2}{*}{0.94$^{+0.37}_{-0.26}$} & \multirow{2}{*}{$107^{+42}_{-29}$ } & \multirow{2}{*}{ 0.878$^{+0.488}_{-0.340}$} & \multirow{2}{*}{$7853.4^{+2.1}_{-2.2}$} & \multirow{2}{*}{$88.5^{+0.7}_{-0.6}$} & \cite{MCPIII}, \\
& 43:46:56.7465 & & & & & &\cite{MCPVI}\\ 
\hline
\multirow{2}{*}{UGC 3789} & 07:19:30.9490 & \multirow{2}{*}{1.16 $\pm$ 0.12} & \multirow{2}{*}{49.6 $\pm$ 5.1 } & \multirow{2}{*}{2.338 $\pm$ 0.341} & \multirow{2}{*}{3320 $\pm$ 1} & \multirow{2}{*}{90.6 $\pm$ 0.4} & \cite{MCPIII}, \\
& 59:21:18.3150 & & & & & & \cite{MCPIV}\\ 
\hline
\multirow{2}{*}{CGCG 074-064} & 14:03:04.457746 & \multirow{2}{*}{$2.42^{+0.22}_{-0.20}$} & \multirow{2}{*}{$87.6^{+7.9}_{-7.2}$ } & \multirow{2}{*}{2.762$^{+0.353}_{-0.321}$} & \multirow{2}{*}{$6908.9^{+1.8}_{-1.9}$} & \multirow{2}{*}{90.8 $\pm$ 0.6 } & \multirow{2}{*}{\cite{MCPXI}} \\
& 08:56:51.03483 & & & & & &\\ 
\hline
\multirow{2}{*}{ESO 558-G009} & 07:04:21.0113 & \multirow{2}{*}{1.7 $\pm$ 0.1} & \multirow{2}{*}{107.6 $\pm$ 5.9} & \multirow{2}{*}{1.579$\pm$ 0.127} & \multirow{2}{*}{7595.9 $\pm$ 14} & \multirow{2}{*}{98 $\pm$ 1} & \multirow{2}{*}{\cite{MCPIX}} \\
& -21:35:18.948 & & & & & &\\ 
\hline
\multirow{2}{*}{NGC 2960} & 09:40:36.38370 & \multirow{2}{*}{1.16 $\pm$ 0.05}  & \multirow{2}{*}{$72.2^a$} & \multirow{2}{*}{1.606} & \multirow{2}{*}{4945 $\pm$ 15} &  \multirow{2}{*}{$89^a$ } &  \multirow{2}{*}{\cite{MCPIII}}\\
& 03:34:37.2915 & & & & & &\\ 
\hline
\multirow{2}{*}{NGC 6264} & 16:57:16.12780 & \multirow{2}{*}{3.09 $\pm$ 0.42} & \multirow{2}{*}{144 $\pm$ 19} & \multirow{2}{*}{2.145 $\pm$ 0.406} & \multirow{2}{*}{10189 $\pm$ 1.2} & \multirow{2}{*}{89.5 $\pm$ 0.9} & \cite{MCPIII}, \\
& 27:50:58.5774 & & & & & & \cite{MCPV}\\ 
\hline
\multirow{2}{*}{J0437+2456} & 04:37:03.6840 & \multirow{2}{*}{0.29 $\pm$ .03} & \multirow{2}{*}{65.3 $\pm$ 3.6} & \multirow{2}{*}{ 0.444 $\pm$ 0.052} & \multirow{2}{*}{4809.5 $\pm$ 10.5} & \multirow{2}{*}{81 $\pm$ 1} & \multirow{2}{*}{\cite{MCPIX}} \\
& 24:56:06.837 & & & & & &\\ 
\hline
\multirow{2}{*}{NGC 4388} & 12:25:46.77914 & \multirow{2}{*}{0.84 $\pm$ 0.02} & \multirow{2}{*}{$19^a$} & \multirow{2}{*}{4.421} & \multirow{2}{*}{2527 $\pm$ 1} &  \multirow{2}{*}{-} &  \multirow{2}{*}{\cite{MCPIII}}\\
& 12:39:43.7516 & & & & & &\\
\hline
\multirow{2}{*}{NGC 2273} & 06:50:08.65620 & \multirow{2}{*}{0.75 $\pm$ 0.04}  & \multirow{2}{*}{$25.7^a$} & \multirow{2}{*}{ 2.918} & \multirow{2}{*}{1832 $\pm$ 15} & \multirow{2}{*}{$84^a$ } &  \multirow{2}{*}{\cite{MCPIII}}\\
& 60:50:44.8979 & & & & & &\\
\end{tabular}
\label{tab:prev}
\end{center}
\end{table*}

\section{Statistical fit}
\label{sec:bayes}

Since the observations show that the maser features have circular orbits on the equatorial plane, we can apply our general relativistic model using a Bayesian approach. 

Considering the maser features to be scattered a small angle $\delta \varphi$ about the midline, and the disk inclination to be specified by the polar angle $\theta_0$, we can write the $\chi^2$ of the general relativistic model as follows \citep{Herrnstein05,Nucamendi21}

\begin{equation}
    \chi^2 = \sum_i \frac{\left[ Z_{obs,i} - (1 + Z_{grav} + \epsilon \sin \theta_0 Z_{kin\pm})(1+Z_{rec}) +1 \right]^2}{\sigma^2_{Z_i} + \kappa ^2 \epsilon^2 \sin^2 \theta_0 Z_{kin\pm}^2(1+Z_{rec})^2 } , 
\end{equation}
where $Z_{obs,i}$ represents the measured redshift of the $i$-th maser feature and the rest of the numerator is the total redshift of our model. The quantities $\epsilon$ and $\kappa$ account for an additional bias caused by the scattering angle on the high-redshifted maser features \citep{Herrnstein05}
\begin{align*}
    \epsilon & \approx 1 - \frac{\delta \varphi^2}{2} + \frac{\delta \varphi^4}{24}, & \kappa^2 & \approx \frac{\delta \varphi^4}{4}.
\end{align*}

The term $\sigma_{Z_i}$ in the denominator represents the error of the total redshift, i.e. the variation $\delta Z_{tot1,2}$, which reads
\begin{align}
    \delta Z_{tot1,2} = (\delta Z_{grav} \pm \delta Z_{kin})(1+Z_{rec}),
\end{align}
where the errors of the gravitational redshift $\delta Z_{grav}$ and the kinematic redshift $\delta Z_{kin}$ are
\begin{align}
    \delta Z_{grav} &= - \frac{3}{2} (1+Z_{grav})^3 \left( \frac{M}{r_e} \right) \left(\frac{\delta r_e}{r_e} \right) , &\\
    \delta Z_{kin} &= \frac{1}{2} \epsilon \sin \theta_0 Z_{kin\pm}^3 \left(\frac{6M^2-r_e^2}{Mr_e}\right) \left(\frac{\delta r_e}{r_e}\right).
\end{align}

Finally, $\delta r_e$ is the error associated with the rotated position $(x, y)$ of each maser feature

\begin{align}
    \delta r_e = \sqrt{\left(\frac{x-x_0}{r}\right)^2 \sigma_{x}^2 + \left(\frac{y-y_0}{r}\right)^2 \sigma_{y}^2 }.
\end{align}

Since we rotated the maser disk by the position angle $p_0$, we propagate the error in the $x$ offset $\sigma _x$ and the $y$ offset $\sigma _y$ as follows
\begin{align}
   \label{eq:errx}
   \sigma _x^2 =& \left(\frac{\partial x}{\partial x'} \sigma _{x'}\right)^2 + \left(\frac{\partial x}{\partial y'} \sigma _{y'}\right)^2 + \left(\frac{\partial x}{\partial p_0} \sigma _{p_0}\right)^2 , & \\
   \label{eq:erry}
   \sigma _y^2 = & \left(\frac{\partial y}{\partial x'} \sigma _{x'}\right)^2 + \left(\frac{\partial y}{\partial y'} \sigma _{y'}\right)^2 + \left(\frac{\partial y}{\partial p_0} \sigma _{p_0}\right)^2 ,
\end{align}
where $\sigma _{x'}$ and $\sigma _{y'}$  are the reported observational errors of the Right Ascension ($x'$) and Declination ($y'$), respectively. The term $\sigma _{p_0}$ accounts for the uncertainty in the position angle; however, we can obviate this term in the propagation since the changes produced on the respective $M/D$ values (and their corresponding errors) when taking it into account are negligible with respect to the estimation without this term (less than $0.06\%$, which is well into the $M/D$ error).

\subsection{Parameter fitting: priors and posteriors}

The BH parameters that undergo a Bayesian statistical fit are the mass-to-distance ratio $M/D$, the recessional redshift $Z_{rec}$, the horizontal offset $x_0$ and the vertical offset $y_0$ of the BH position.

On the other hand, for half of our set of galaxies, the $y_0$ offset can be statistically estimated and for the other half it cannot. For the latter set of galaxies, $y_0$ was fixed at the geometric center of the disk formed by the redshifted and blueshifted maser features, estimating only three parameters for them. Moreover, for this set the $y$ offset is an irrelevant parameter in the sense that it leads to insignificant changes in the $M/D$ best values. These changes are indicated for the galaxies with only three fitted parameters in what follows.

\begin{figure*}
	\includegraphics[scale=0.8]{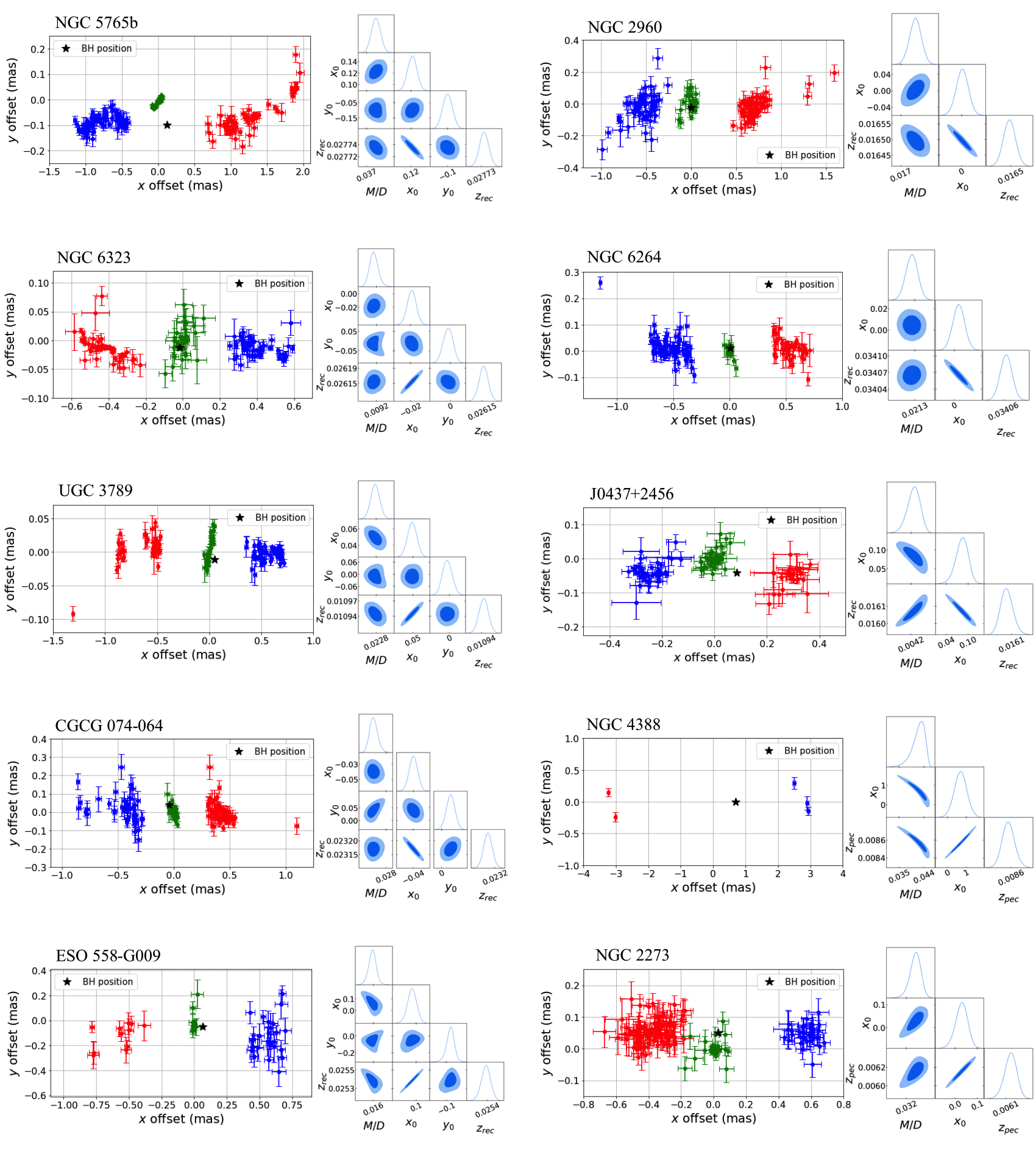}
    \caption{Maser disks and posterior probability distributions. The maser disk plots show the three groups of maser features viewed edge on with their observational error. The star indicates the best fit for the BH position on the disk. The left column shows the galaxies with 4 fitted parameters and the right column the galaxies with three parameters fitted. The blue graphs show the posterior probability distribution with the contour levels corresponding to 1$\sigma$ and 2$\sigma$ confidence regions.}
    \label{fig:posteriors}
\end{figure*}

Finally, for each galaxy we fixed the scattering angle of maser features $\delta \varphi$ such that the reduced $\chi^2$ reaches a value close to unity. The posterior fitting results are summarized in Table \ref{tab:Results4par} and the posterior probability distribution is displayed in Fig. \ref{fig:posteriors} along with the maser feature distributions.

\begin{table*}
\begin{center}
\caption{Posterior values of the fit. Column 1: Name of the source. Column 2: Fitted mass-to-distance ratio. Column 3: Horizontal offset ($x_0$) of the BH found by the Bayesian fit. Column 4: Vertical offset ($y_0$) of the BH found by the Bayesian fit. From NGC 5765b to ESO 558-G009 the $y_0$ offset is found by the statistical fit. For NGC 2960 to NGC 2273 the $y_0$ offset is fixed at the geometric center of the disk formed by the high-redshifted maser features. Column 5: Recessional redshift found by the fit. The peculiar redshift parameter $Z_{pec}$ is fitted for the NGC 2273 and 4388 galaxies. Column 6: Recession velocity defined after the optical definition of the redshift, i.e., $v_{rec} = cZ_{rec}$. Column 7: Scattering angle in which the maser features are spread on the azimuth angle about the midline. Column 8: Reduced $\chi^2$ of the best fit.}
\begin{tabular}{ c c c c c c c c}
Source & $M/D$ & $x_0$ & $y_0$ & $Z_{rec}$ & $v_{rec}$ & $\delta \varphi$ & $\chi^2_{red}$ \\
& ($10^5 M_{\odot}/$Mpc) & (mas) & (mas) & ($10^{-2}$) & (km/s) & (°)&\\
\hline 
\hline 
\multirow{2}{*}{NGC 5765b} & \multirow{2}{*}{$3.727 \pm0.013$} & \multirow{2}{*}{$0.122^{+0.008}_{-0.009}$} & \multirow{2}{*}{$-0.099^{+0.034}_{-0.033}$} & \multirow{2}{*}{$2.773 \pm 0.0008$} & \multirow{2}{*}{$8314.67^{+2.43}_{-2.38}$}& \multirow{2}{*}{10} &  \multirow{2}{*}{1.165} \\
& & & & & &\\
\hline
\multirow{2}{*}{NGC 6323} & \multirow{2}{*}{$0.916\pm 0.005$} & \multirow{2}{*}{$-0.018 \pm0.007$} & \multirow{2}{*}{$-0.012^{+0.026}_{-0.027}$} & \multirow{2}{*}{$2.615 \pm0.001$} & \multirow{2}{*}{$7840.78^{+3.73}_{-3.71}$}& \multirow{2}{*}{11} &  \multirow{2}{*}{1.276} \\
& & & & & &\\
\hline 
\multirow{2}{*}{UGC 3789} & \multirow{2}{*}{$2.277\pm 0.009$} & \multirow{2}{*}{$0.049\pm0.006$} & \multirow{2}{*}{$-0.011^{+0.026}_{-0.025}$} & \multirow{2}{*}{$1.094\pm0.001$} & \multirow{2}{*}{$3281.07^{+2.89}_{-2.91}$}& \multirow{2}{*}{10} &  \multirow{2}{*}{1.430} \\
& & & & & &\\
\hline 
\multirow{2}{*}{CGCG 074-064} & \multirow{2}{*}{$2.749^{+0.017}_{-0.015}$} & \multirow{2}{*}{$-0.039\pm-0.006$} & \multirow{2}{*}{$0.040\pm0.019$} & \multirow{2}{*}{$2.317\pm 0.002$} & \multirow{2}{*}{$6946.30^{+5.33}_{-5.30}$}& \multirow{2}{*}{11} &  \multirow{2}{*}{1.289} \\
& & & & & &\\
\hline 
\multirow{2}{*}{ESO 558-G009} & \multirow{2}{*}{$1.594^{+0.025}_{-0.026}$} & \multirow{2}{*}{$0.064^{+0.041}_{-0.042}$} & \multirow{2}{*}{$-0.050^{+0.045}_{-0.050}$} & \multirow{2}{*}{$2.538\pm 0.005$} & \multirow{2}{*}{$7610.05^{+15.48}_{-15.70}$}& \multirow{2}{*}{7} &  \multirow{2}{*}{1.188} \\
& & & & & &\\
\hline \hline
\multirow{2}{*}{NGC 2960} & \multirow{2}{*}{$1.738^{+0.015}_{-0.016}$} & \multirow{2}{*}{$0.002\pm0.016$} & \multirow{2}{*}{-0.020} & \multirow{2}{*}{$1.650\pm0.002$} & \multirow{2}{*}{$4945.30^{+6.22}_{-6.23}$}& \multirow{2}{*}{15} &  \multirow{2}{*}{1.488} \\
& & & & & &\\
\hline 
\multirow{2}{*}{NGC 6264} & \multirow{2}{*}{$2.126\pm0.007$} & \multirow{2}{*}{$0.004\pm0.006$} & \multirow{2}{*}{$0.012$} & \multirow{2}{*}{$3.406\pm 0.001$} & \multirow{2}{*}{$10212.47\pm 3.44$}& \multirow{2}{*}{10} &  \multirow{2}{*}{1.304} \\
& & & & & &\\
\hline 
\multirow{2}{*}{J0437+2456} & \multirow{2}{*}{$0.425^{+0.010}_{-0.012}$} & \multirow{2}{*}{$0.084^{+0.016}_{-0.018}$} & \multirow{2}{*}{$-0.040$} & \multirow{2}{*}{$1.607\pm0.004$} & \multirow{2}{*}{$4817.78^{+11.58}_{-10.78}$}& \multirow{2}{*}{0} &  \multirow{2}{*}{1.038} \\
& & & & & &\\
\hline 
\multirow{2}{*}{NGC 4388} & \multirow{2}{*}{$4.233^{+0.185}_{-0.268}$} & \multirow{2}{*}{$0.696^{+0.330}_{-0.334}$} & \multirow{2}{*}{0.000} & \multirow{2}{*}{$0.857\pm0.007$} & \multirow{2}{*}{$2568.33^{+22.15}_{-21.23}$}& \multirow{2}{*}{9} &  \multirow{2}{*}{1.369} \\
& & & & & &\\
\hline 
\multirow{2}{*}{NGC 2273} & \multirow{2}{*}{$3.283 ^{+0.045}_{-0.051}$} & \multirow{2}{*}{$0.029^{+0.026}_{-0.028}$} & \multirow{2}{*}{0.050} & \multirow{2}{*}{$0.616^{+0.006}_{-0.007}$} & \multirow{2}{*}{$1845.81^{+19.42}_{-21.03}$}& \multirow{2}{*}{15} &  \multirow{2}{*}{1.627} \\
& & & & & &\\
\end{tabular}
\label{tab:Results4par}
\end{center}
\end{table*}

\subsubsection{NGC 5765b}

We found the scattering angle $\delta \varphi = 10$° to be suitable for this system and performed the Bayesian statistical fit with four parameters rendering a reduced $\chi ^2$ of 1.165. Since the disk of this galaxy has an inclination angle of  $\theta_0 = 94.5$° \citep{MCPVIII}, the LOS masers seem misaligned with the redshifted and blueshifted masers. Besides, more redshifted maser features were observed in a wider $y$ offset range in comparison to the blueshifted ones. As a result of the Bayesian statistical fit, we found the BH position to be near the geometric center formed by the redshifted and the blueshifted maser features.

\subsubsection{NGC 6323}

The three groups of maser features in this galaxy are almost aligned with each other, with an inclination angle of nearly 90° \citep{MCPVI}. We fitted four parameters giving a reduced $\chi^2$ of 1.251 with a BH position on the sky of (-0.018 , -0.012) mas with respect to the geometric center formed by the systemic masers; this position locates the BH in the LOS maser features slightly below the origin and aligned with the redshifted and blueshifted masers. The angle of scatter for the highly frequency-shifted maser features is $\delta \varphi = 11$°.

\subsubsection{UGC 3789}

Although this maser disk is the thinnest of our group of megamaser systems, it was possible to estimate the BH offset $y_0$. The statistical fit rendered the BH position to be near the geometric center, with the maser features scattered about the midline with an angle $\delta \varphi = 10 \degr$. The best reduced $\chi^2$ fit was 1.430.

\subsubsection{CGCG 074-064}

Almost all of the maser features are aligned with each other, we see the disk practically edge on, with an inclination angle of $90.8\degr$ \citep{MCPXI}. For this galaxy we fitted four parameters with a scattering angle of $\delta \varphi = 10\degr$, resulting the BH position to be near the center of the LOS maser features, with a reduced $\chi^2 = 1.600$.

\subsubsection{ESO 558-G009}

The maser disk of this galaxy has an inclination angle of $\theta_0 = 98$° \citep{MCPIX}, for which the LOS maser features are in an upper $y$ offset than the high-redshifted masers. The statistical Bayesian fit with four parameters found the BH position to be aligned with the LOS masers in the $x$ offset and below the origin in the $y$ offset. The scattering angle for the redshifted and blueshifted maser features in the azimuthal angle was $\delta \varphi = 7$°, giving a reduced $\chi^2 $ of 1.188.

\subsubsection{NGC 2960}

The observations show that the maser features of this galaxy are horizontally aligned; nevertheless, it was not possible to statistically estimate the BH $y_0$ offset. By varying the $y_0$ position along the height of the disk, we found variations on $M/D$ up to $4\%$ with respect to the $M/D$ ratio found with $y_0$ fixed at the origin (the geometric center of the systemic masers). By fitting three parameters in this galaxy, we obtained a $x_0$ position only 0.004 mas away from the origin with a reduced $\chi^2$ of 1.511 and a scattering angle $\delta \varphi = 15$°.

\subsubsection{NGC 6264}

We performed the Bayesian statistical fit for three parameters, fixing $y_0$ in the geometric center of the disk formed by the highly frequency-shifted masers at a $y$ offset of 0.030 mas. Similarly as for NGC 2960, we varied the $y_0$ offset  along the height of the disk and found variations on $M/D$ of less than $0.3 \%$, inside the error. The angle on which the highly frequency-shifted masers are spread about the midline is $\delta \varphi = 10$°, rendering a reduced $\chi^2 = 1.301$.

\subsubsection{J0437+2456}

The inclination angle for this maser disk is of $\theta_0 = 81$°, hence the three groups of masers do not seem horizontally aligned. The $y_0$ offset was fixed at the geometric center of the highly frequency-shifted masers $y_0 = -0.040$ mas. By varying the $y_0$ position along the height of the maser disk we found variations in $M/D$ up to $\sim 2\%$, which is inside the error of the posterior $M/D$. For this megamaser system there is no need to consider a spread in the azimuthal angle since the statistical fit with $\delta \varphi =0\degr$ already renders a reduced $\chi^2 = 1.038$.

\subsubsection{NGC 4388}
For this megamaser system only five maser features were observed: two redshifted and three blueshifted. Due to the lack of systemic maser observations, the authors \citet{MCPIII} do not estimate the inclination angle; therefore, we set $\theta_0 = 90\degr$ and perform the statistical fit for three parameters. Additionally, (according to \citeauthor{MCPIII}) this galaxy is located at 19  Mpc, meaning that the galaxy is not into the Hubble flow. For this reason, we use $Z_{boost}$ instead of the $Z_{rec}$ composition in Eq. (\ref{eq:ztot}) and fit for the peculiar redshift $Z_{pec}$. The best scattering angle we found was $\delta \varphi = 9\degr$ and a reduced $\chi^2 = 1.369$. We fixed the $y$ offset of the BH at $y_0 = 0.0$ mas and varied this position along the height of the disk. The variations in $M/D$ were up to $0.48 \%$, which is below the error of the $M/D$ posterior value.

\subsubsection{NGC 2273}
Similarly as NGC 4388, this galaxy is not into the Hubble flow since it is 25 Mpc away \citep{MCPIII}. We therefore fit for the peculiar redshift $Z_{pec}$ as explained above. We fixed the BH $y$ offset at $y_0 = 0.050$ mas and, varying $y_0$ along the disk we found differences in $M/D$ up to $0.6\%$, which is less than the percentage error of the posterior $M/D$. The obtained reduced $\chi^2$ is of 1.627 for this astrophysical system with a scattering angle of $\delta \varphi = 15$°.\\

Within our approach we can estimate just the mass-to-distance ratio since we do not have information about the distance to the BH source obtained in an independent experiment. The fits of the distance to the  BHs in the MCP reports make use of the same frequency shift data that we use for the $M/D$ statistical fit. 
Thus, we can not compare the masses of our BHs with the masses reported in the MCP works. However, we can compare the mass-to-distance ratio of both studies by dividing the mass over the distance of the MCP results and propagating accordingly their corresponding uncertainties, as it was done in Table \ref{tab:prev}. Notwithstanding, comparison of the uncertainties of these most probable values is misleading since the uncertainty in the mass-to-distance ratio of the MCP results inherits the large distance error during propagation.

In order for our general relativistic method to make a meaningful improvement on independent mass and distance estimations, compared to the published MCP results, we need to find the way to decouple the BH mass and the distance to the BH in the model for the studied astrophysical systems. This is a work which is currently in progress.

\begin{table}
\begin{center}
\caption{Gravitational redshift of the masers closest to the black hole.}
\begin{tabular}{ c c c c}
& Distance to the & Gravitational & Associated\\
Source & closest maser& redshift & velocity\\
& (mas) & ($10^{-6}$) & (km/s)\\
\hline \hline
NGC 5765b & 0.537 & 10.259 & 3.075 \\
NGC 6323 & 0.215 & 6.310 & 1.891 \\ 
UGC 3789 & 0.305 & 11.042 & 3.310 \\ 
CGCG 074-064 & 0.252 & 16.098 & 4.826 \\ 
ESO 558-G009 & 0.380 & 6.199 & 1.858 \\ 
NGC 2960 & 0.291  & 8.830 & 2.647 \\ 
NGC 6264 & 0.336 & 9.357 & 2.805 \\
J0437+2456 & 0.140 & 4.476 & 1.341 \\ 
NGC 4388 & 1.829 & 3.422 & 1.026 \\ 
NGC 2273 & 0.211 & 22.927 & 6.873 \\ 
\end{tabular}
\label{tab:zgrav}
\end{center}
\end{table}

\section{Discussion and conclusions}
\label{sec:conclusions}

Using a general relativistic formalism, we found the expression for the redshift of photons emitted by a test particle moving in the Schwarzschild metric. The total redshift comprises the gravitational redshift, the kinematic frequency shift and the recessional redshift, the former being a relativistic effect with no Newtonian analog. 

By fitting the observations with our model, we were able to estimate the mass-to-distance ratio $M/D$, the position $(x_0, y_0)$ and the recessional redshift $Z_{rec}$ of ten BHs using spectroscopy and astrometry data of the maser features on their accretion disks. Considering the distances listed in Table \ref{tab:prev} we can compute the BH masses of each galaxy, obtaining masses within the range $10^6-10^7M_{\sun}$. This fact indicates that every studied BH is a supermassive BH, in agreement with the hypothesis of existence of supermassive BHs at AGN cores \citep{Rees84,Kormendy13}.

For all the analysed galaxies, the BH position ($x_0, y_0$) appears within the region of the systemic masers, which is physically in concordance with the maser emission mechanism. Given the lack of significant change in the $M/D$ posterior values when varying the $y_0$ offset, we conclude that there is no sensitivity for the estimation of this parameter in the set of five galaxies where only three parameters could be fitted.

On the other hand, for the galaxies located into the Hubble flow (>30 Mpc), the posterior recession velocities found in the statistical fit $v_{rec}$ increase as $D$ does due to the accelerated expansion of the Universe \citep{Riess98,Planck15}. For these megamaser systems the peculiar and cosmological redshifts are composed into the corresponding recessional one in concordance with the explanation given at the end of Sec. \ref{sec:GR}.

We calculated the gravitational redshift for the closest maser to the BH in each galaxy in Table \ref{tab:zgrav} obtaining values between $3.422 - 22.927 \times 10^{-6}$ with the corresponding velocities $1.03 - 6.87$ km/s. The strongest gravitational redshift belongs to NGC 2273, where the accretion disk is the smallest one and therefore there is a closer proximity of the maser features to the BH, leading to a stronger redshift. However, the values of this redshift are generally smaller than the errors of the corresponding total redshifts, meaning that a higher precision in frequency shift measurements is needed in order to unambiguously distinguish the gravitational redshift from the kinematic and recessional ones. Nevertheless, the gravitational redshift detection could be improved if in a future analysis we take into account the redshift of the systemic masers since they present a low magnitude kinematic redshift, giving rise to an approximately isolated combination of the gravitational and the recessional redshifts.

\section*{Acknowledgements}

The authors thank the Megamaser Cosmology Project researchers for making the observational data used in this work publicly available. All authors are grateful to A. Villalobos-Ramírez and O. Gallardo-Rivera for fruitful discussions and to FORDECYT-PRONACES-CONACYT for support under grant No. CF-MG-2558591; U.N. also was supported under grant CF-140630. A.H.-A. and U.N. thank SNI and PROMEP-SEP and were supported by grants VIEP-BUAP No. 122 and CIC-UMSNH, respectively. D.V. acknowledges financial assistance from CONACYT through the grant No. 1071008.

\section*{Data Availability}

The data used in this work belongs to their corresponding authors and are available at the following references: \citet{MCPIII} (NGC 2273, NGC 2960, NGC 4388, NGC 6264 and NGC 6323), \citet{MCPIV} (UGC 3789), \citet{MCPVIII} (NGC 5765b), \citet{MCPIX} (ESO 558-G009 and J0437+2456), and \citet{MCPXI} (CGCG 074-064).



\bibliographystyle{mnras}
\bibliography{bib} 





\bsp	
\label{lastpage}
\end{document}